\documentclass[aps,twocolumn,showpacs,preprintnumbers,amsmath,amssymb,floatfix]{revtex4}

\usepackage{graphicx}
\usepackage{dcolumn}
\usepackage{bm}
\usepackage[german,american]{babel}

\begin{document}


\title{Nonlinear Ionic Conductivity of Thin Solid Electrolyte Samples:
Comparison between Theory and Experiment}

\author{Andreas Heuer, Sevi Murugavel, Bernhard Roling}

\affiliation{Institut f\"{u}r Physikalische Chemie and
Sonderforschungsbereich 458, \\
Westf\"{a}lische Wilhelms-Universit\"{a}t M\"{u}nster, \\
Corrensstr. 30, 48149 M\"{u}nster, Germany}

\date{\today}

\begin{abstract}
Nonlinear conductivity effects are studied experimentally and
theoretically for thin samples of disordered ionic conductors.
Following previous work in this field the {\it experimental
nonlinear conductivity} of sodium ion conducting glasses is analyzed 
in terms of apparent hopping distances. Values up to 43 \AA \, are obtained. 
Due to higher-order
harmonic current density detection, any undesired effects arising
from Joule heating can be excluded. Additionally, the influence of
temperature and sample thickness on the nonlinearity is explored.
From the {\it theoretical side} the nonlinear conductivity in a
disordered hopping model is analyzed numerically. For the 1D case
the nonlinearity can be even handled analytically. Surprisingly,
for this model the apparent hopping distance scales with the
system size. This result shows that in general the
nonlinear conductivity cannot be interpreted in terms
of apparent hopping distances. Possible extensions of the model
are discussed.

\end{abstract}

\pacs{66.10.Ed, 66.30.Hs, 61.43.Fs} 
\maketitle

\section{\label{introduction}Introduction}

One important method for improving the properties of solid ionic conductors
with regard to applications in microbatteries, fuel cells and electrochromic
devices is the preparation of thin films \cite{Noda04, Appetecchi03,
Jonghe03, Lu02, Rowley02, Tarascon01, Murray99, Croce98}.
Both the electrical resistance and the
mechanical stiffness of the film decrease with decreasing thickness.
The decrease of the electrical resistance is important for improving
the power density of microbatteries, the efficiency of fuel
cells and the switching time of electrochromic devices. A low mechanical
stiffness leads to a better processibility of the film.

For the integration of a thin film into a lithium microbattery, an electrochemical
window in which the film is stable from 0-5 V is required \cite{Tarascon01,
Noda04}. The application of 5~V to samples with
thicknesses of 100 nm and less leads to high electric field strengths in the
samples, which result in field-dependent ion transport effects.
Generally, the ionic conductivity of solid electrolytes increases with increasing
field strength \cite{Hyde86, Barton96, Isard96, Tajitsu96, Tajitsu98}. Thus,
field-dependent ion transport is of potential interest for improving the
applicability of thin-film electrolytes.

Mathematically, the field-dependent electrical properties of thin solid
electrolyte samples can be described by:
\begin{equation}
j_{dc}(E_{dc})= \sigma_{1,dc}\;E_{dc} + \sigma_{3,dc}\;E^3_{dc}
+ \sigma_{5,dc}\;E^5_{dc} + ...
\end{equation}
Here, the linear dependence of the dc current density $j_{dc}$
on the dc electric field $E_{dc}$ at low field strengths
is characterised by the low-field dc conductivity $\sigma_{1,dc}$,
while the nonlinear dependence at high field strengths is
characterised by the higher-order conductivity
coefficients $\sigma_{3,dc}$, $\sigma_{5,dc}$ etc.

A first step to interpret the nonlinearity of the conductivity
is to consider a simple regular hopping model with distance $a$ between
adjacent sites. For this model it turns out that
\begin{equation}
\label{eq:sinh}
j_{dc}(E_{dc}) \propto \sinh(q \beta a E_{dc}/2)
\end{equation}
with $\beta = 1/k_B T$. Thus, in the framework of this model, it
is possible to extract the hopping distance $a$ from
field-dependent electrical data. Although the model is too simple
to provide a realistic description of ion conduction in disordered
materials, such as glasses and polymer electrolytes, the
$j_{dc}(E_{dc}$) curves of many real ion conductors can be
reasonably well fitted by Eq. (\ref{eq:sinh}). However, the values
obtained for the hopping distance $a$ are much larger than typical
distances between neighboring sites in ionic conductors. For
instance, in the case of ion conducting glasses, values in the
range from  15 \AA$\;$ to 30 \AA$\;$ have been found
\cite{Hyde86, Barton96, Isard96}.

These experimental results are based on measurements using dc electric field. 
A major drawback of
this method is the lack of information about Joule heating effects. Joule heating may
lead to an increase of the sample temperature, resulting in an increase of
the ionic conductivity. In contrast, the application of ac electric fields
allows for a direct
differentiation between nonlinear ion transport and Joule heating. Nonlinear
ion transport leads to higher harmonic contributions to the current density
spectrum, while this is not the case for Joule heating \cite{Murugavel04}.
The field dependence of the higher harmonic contributions can be used to
determine values for the higher order--conductivity coefficients $\sigma_{3,dc}$,
$\sigma_{5,dc}$ etc. \cite{Murugavel04}. Thus, by means of nonlinear
ac conductivity spectroscopy it is possible to differentiate between the
higher-order conductivity coefficients, while the application of dc
fields yields only one $j_{dc}(E_{dc}$) curve.

The values $\sigma_{1,dc}$ and $\sigma_{3,dc}$ can be used to define
an apparent hopping distance:
\begin{equation}
\label{aappdef}
 a_{app}^2 \equiv 24 \sigma_{3,dc} /(\sigma_{1,dc}\,q^2\,\beta^2)
\end{equation}
This definition of $a_{app}$ implies that for the regular hopping
model, the simple relation  $a_{app} = a$ holds. However, two of
the present authors have shown that for different sodium ion
conducting glasses, the apparent hopping distances $a_{app}$ are
in a range from 39\AA$\;$ to 55\AA$\;$ \cite{Murugavel04}. Thus, the
calculation of $a_{app}$ by means of Eq. (\ref{aappdef}) leads to
higher values than fits of  $j_{dc}(E_{dc}$) data by means of Eq.
(\ref{eq:sinh}). This implies that Eq. (\ref{eq:sinh}) does not
provide an exact description of the nonlinear conductivity of
ionic conductors. This is confirmed by the observation of negative
values for $\sigma_{5,dc}$ \cite{Murugavel04}, whereas the
validity of Eq. (\ref{eq:sinh}) implies positive values for
$\sigma_{5,dc}$.

In this paper, we rationalize these experimental results by considering
simple hopping models with strong site and barrier disorder. We find that
in the framework of such models, the apparent hopping distance
$a_{app}$ is, indeed, considerably larger than the distance $a$ between adjacent
sites. However, $a_{app}$ shows an unexpected dependence on the
thickness of the model systems in the direction of the applied electric
field. In order to check whether also real ionic conductors exhibit
such a thickness dependence of their nonlinear electrical properties,
we have carried out nonlinear ac conductivity measurements on
ion conducting glass samples of different thickness. We compare
theoretical and experimental results, and we discuss implications of
these results for further theoretical and experimental work.

\section{\label{Theory}Theory}

\begin{figure}
\includegraphics[width=8.6cm]{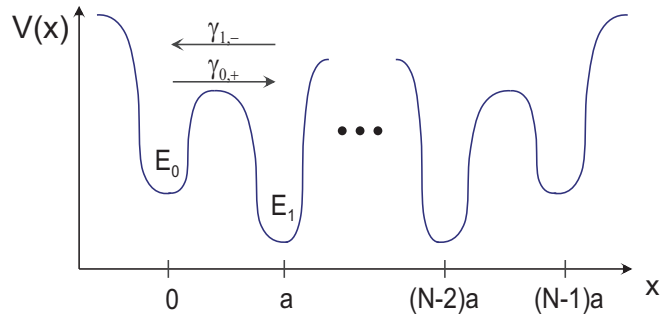}
\caption{\label{sketch} Sketch of the 1D hopping model, used in
the theoretical analysis. Periodic boundary conditions are employed.
The model is point symmetric.}
\end{figure}

First, we consider a general 1D-hopping model with site and
barrier disorder; see Fig. \ref{sketch}. The individual $N$ sites
are characterized by energies $E_i$. With an external field the
hopping rates for leaving site $i$ are $\Gamma_{i,\pm}$ where the
sign denotes whether the hopping is along (+) or opposite (-) to
the field. The field dependence of $\Gamma_{i,\pm}$ can be
expressed as $\Gamma_{i,\pm}(u) = \exp(\pm u) \gamma_{i,\pm}$ with
\begin{equation}
u = q \beta a E_{dc}/2.
\end{equation}
Correspondingly, $\gamma_{i,\pm}$ are the hopping rates without
applied field. Detailed balance requires
$\gamma_{i,+}/\gamma_{i+1,-} = \exp(-\beta(E_{i+1} - E_i))$.  To
get a stationary long-time solution for the current, periodic
boundary conditions are used. Physically, this reflects the
interaction of both electrodes via the voltage source.

The time evolution of the {\it single-particle} problem is
governed by the rate equations
\begin{eqnarray}
(d/dt) p_i(t) && =  - \Gamma_{i,+}(u) p_i(t) - \Gamma_{i,-}(u)
p_i(t)\nonumber \\ & &  + \Gamma_{i-1,+}(u) p_{i-1}(t) +
\Gamma_{i+1,-}(u) p_{i+1}(t).
\end{eqnarray}
The stationary long-time solutions are denoted as $p_i^\infty(u)$.
The argument expresses their dependence on the electric field.
Finally the current $j_{dc}(u)$ in the long-time limit can be
written as
\begin{equation}
\label{current} j_{dc}(u)/(qa) \equiv \tilde{j}(u) =
p_{i}^\infty(u) \Gamma_{i,+}(u) - p_{i+1}^\infty(u)
\Gamma_{i+1,-}(u).
\end{equation}
Note that the current between all pairs of sites is identical.
Therefore $\tilde{j}(u)$ and thus $j_{dc}(u)$ do not depend on the
index $i$.

This property can be used to solve the system of equations
analytically for arbitrary $N$; see Refs.
\cite{Ambegaokar69,Derrida83,Bouchaud90,Kehr97} for detailed
information.

Using a notation which is appropriate for our further analysis the
solution can be written as
\begin{equation}
\label{jdef} \tilde{j}(u) = (\exp(Nu) - \exp(-Nu))/B
\end{equation}
with
\begin{equation}
\label{bdef}
 B = \sum_{m=0}^{N-1} A_m \exp((1-N+2m)u).
\end{equation}
The disorder is contained in
\begin{equation}
A_m = \sum_{i=0}^{N-1}f(m,i)
\end{equation}
with
\begin{equation}
\label{deff}
 f(m,i) \equiv  \exp(-\beta(E_{i+m} -
E_{i}))/\gamma_{i,-}.
\end{equation}
In case that the
disorder of adjacent sites is uncorrelated one obtains $\langle
A_1 \rangle = ... = \langle A_{N-2} \rangle \equiv b_1$ and
$\langle A_0 \rangle = \langle A_{N-1} \rangle = b_0$, where the
brackets denote the disorder average. Please note that $\langle
A_m \rangle $ as a sum of $N$ terms trivially scales like $N^1$.

In what follows we consider the situation that the energy
landscape is point symmetric. In this way one can guarantee that
in analogy to the experimental situation the current only depends
on odd powers of the electric field. In particular one strictly
has $A_m = A_{N-1-m}$.

Here we are interested in a Taylor-expansion of $\tilde{j}(u)$
with respect to $u$. Expansion of the numerator and the denominator
yields
\begin{equation}
\tilde{j}(u) =  \frac{2Nu + N^3u^3/3 + ...}{\sum_m A_m + (1/2)
\sum_m A_m (-N+1+2m)^2 u^2 + ...}.
\end{equation}
Reordering of terms yields up to order $u^2$
\begin{equation}
\label{jx20}
 \tilde{j}(u) =  \frac{2Nu}{\sum A_m} \left [1 + u^2
\left ( N^2/6 - \frac{\sum_m A_m (-N+1+2m)^2}{2\sum_m A_m }\right
) \right ].
\end{equation}
From this the low-field conductivity $\sigma_{1,dc}$ can be
written as:
\begin{equation}
\frac{\sigma_{1,dc}}{q^2 \beta a^2} = \left \langle \frac{N}{\sum_m A_m}
\right \rangle.
\end{equation}
The brackets denote the average over the disorder.

In a first step we use the property that in the thermodynamic
limit $N \rightarrow \infty$ one has
\begin{equation}
\label{disorder}
 \left \langle \frac{1}{\sum_m A_m} \right \rangle =
\frac{1}{\langle \sum_m A_m \rangle}.
\end{equation}
This relation is derived in the Appendix.

With the same arguments one can replace in the disorder average of
$\tilde{j}(u)$ in Eq. (\ref{jx20}) all terms $\sum_m A_m$ by
$\langle \sum_m A_m \rangle$. Combining Eq. (\ref{aappdef})  with
Eq. (\ref{jx20}) and using this preaveraging of $\sum_m A_m$ one
obtains in the limit of large $N$ the following relation for
$a^2_{app}$:
\begin{equation}
 a^2_{app}/a^2 = N^2 -\frac{3 \sum_m \langle A_m
(-N+1+2m)^2\rangle }{\langle \sum_m A_m \rangle}.
\end{equation}

Straightforward calculation yields
\begin{equation}
\langle \sum_m A_m \rangle = N b_1 + 2 (b_0 - b_1)
\end{equation}
and
\begin{eqnarray}
\nonumber
\langle \sum_m  A_m (-N+1+2m)^2\rangle &=& (1/3) b_1 (N^3 - N) \\
&+& 2 (b_0 - b_1) (N-1)^2
\end{eqnarray}
Expansion of the denominator with respect to $1/N$ finally yields
\begin{equation}
\label{asympt}
 a^2_{app}/a^2 = 1 + 4 N (b_1 - b_0)/b_1.
\end{equation}
This shows that for large $N$ one finds the scaling relation
$a^2_{app} \propto N$.

What is the origin of the $N$-dependence? In the case of vanishing
disorder one has $p_i^\infty (u)= 1/N$ for all $u$. Thus the
$u$-dependence of the current $\tilde{j}(u)$, see Eq.
(\ref{current}), exclusively results from the trivial
$u$-dependence of the rates $\Gamma_{i,\pm}(u)$ and gives rise to
$a^2_{app}/a^2 = 1$. In contrast, for disorder one also obtains
contributions from $p_i^\infty (u)$. After solving the N linear
equations, characterizing the stationary case, the expansion of $
p_i^\infty (u)$ with respect to $u$ contains terms proportional to
$N$, which are responsible for the $N$-dependence of
$a^2_{app}/a^2$. For the N-dependence of the $\sigma_{i,dc}$ one
finds $\sigma_{1,dc} \propto N^{-1}$ and $\sigma_{3,dc} \propto
N^0$, thus giving rise to $a^2_{app} \propto
\sigma_{3,dc}/\sigma_{1,dc} \propto N$. The scaling of
$\sigma_{1,dc}$ simply means that the conductivity is proportional
to the particle concentration, i.e. $1/N$. Actually, it is also
possible to calculate $\tilde{j}(u)$ for fixed $u$ in the
thermodynamic limit, thereby revealing non-analytical behavior for
$u=0$. This discussion, however, is beyond the scope of the
present paper.

It may be instructive to calculate $b_0$ and $b_1$ for different
types of disorder. In case of a random barrier model for which all
$E_i$ are constant one obtains $b_0 = b_1$ and thus $a^2_{app}/a^2
= 1$. In case of pure energetic disorder we choose a model for
which $E_i$ is either $e$ or $0$, both with 50\% probability. Then
one finds numerically in the low-temperature limit $u \gg 1$ that
$b_0 \approx 0.24 \exp(\beta e)$ and $b_1 \approx 0.37 \exp(\beta
e)$. For large $N$ one thus gets $a_{exp}^2/a^2 \approx 1.5 N$.

\begin{figure}
\includegraphics[width=8.6cm]{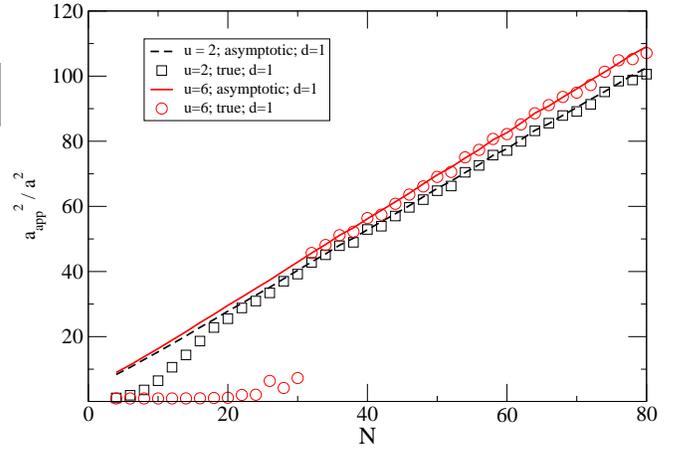}
\caption{\label{results} Comparison of the numerical solution for
$a_{app}$ with the asymptotic analytical solution.}
\end{figure}

In Fig. \ref{results} we show the asymptotic result Eq.
(\ref{asympt}) for $a^2_{app}/a^2$ together with the exact result,
obtained from averaging the current in Eq. (\ref{jx20}) and
finally calculating $a^2_{app}/a^2$ from Eq. (\ref{aappdef}). One
can see that for $N \ge 40$ the large-N limit works very well. For
smaller $N$ the apparent jump length is smaller than expected. We
mention in passing that all data points have been obtained from
averaging over 10000 realisations of the disorder. This large
number is necessary because numerically it turns out that the
standard deviation for the distribution of $a^2_{app}/a^2$ is of
the same order as $a^2_{app}/a^2$ itself (for all $N$). The main
conclusion is that for larger values of $N$ also large values of
$a^2_{app}/a^2$ can be realized. Furthermore we have included data
for $u = 2$ in Fig. \ref{results}. Decrease of $u$ corresponds to
an increase of temperature.  In agreement with the experimental
data (see below) one finds that $a^2_{app}/a^2$ decreases with
increasing temperature. Of course, for $u=0$, one obtains
$a^2_{app}/a^2 = 1$. Actually, as seen from Fig. \ref{results} the
asymptotic result is not strictly linear as naively expected from
Eq. (\ref{asympt}). The reason is that for small $N$ also the
values of $b_0/N$ and $b_1/N$ somewhat depend on $N$ due to
finite-size effects.

So far we have considered a single particle in a disordered
potential. In recent work it has been shown
\cite{Lammert03, Habasaki04, Vogel04a} that it is a more realistic picture
to consider the dynamics of a vacancy rather than a particle. Not
surprisingly, the new set of equations is very similar and the
results for $a^2_{app}/a^2$ turn out to be identical.

\begin{figure}
\includegraphics[width=8.6cm]{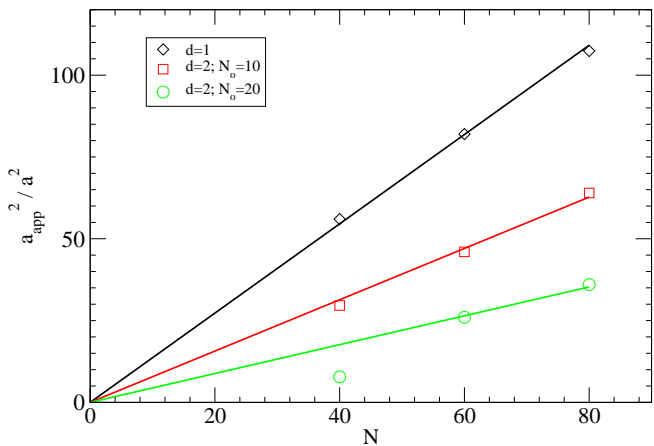}
\caption{\label{data2} Dependence of $a_{app}$ on the size
parallel and orthogonal to the electric field.}
\end{figure}

A more serious deviation from the experimental situation is the
dimensionality. To which degree can the results for the 1D model
be generalized to systems with more than one dimension? To clarify
the influence of the dimension we have extended the model to two
dimensions, also using periodic boundary conditions in the second
dimension. We define $N_o$ as the number of sites orthogonal to
the field. The set of rate equations for the 1D case can be easily
generalized. Solving for the stationary long-time solution is
equivalent to solve a system of linear equations with $N \cdot
N_o$ variables. Unfortunately, it is not possible to solve the
multidimensional case analytically in the way it has been done for
the 1D system. Therefore the solution is purely numerical. After
averaging over the disorder one obtains the results, shown in Fig.
\ref{data2}. Two important conclusions can be drawn from these
results:(i) The scaling $a^2_{app}/a^2 \propto N$ also holds for
two-dimensional systems, albeit only for somewhat larger $N$ (as
can be seen for $N_o = 20$). (ii) The absolute value of
$a^2_{app}/a^2$ strongly decreases with increasing $N_o$.
Comparing $N_o = 10$ and $N_o = 20$ one may speculate that
$a^2_{app}/a^2 \propto N/N_o$.

\section{\label{Experimental}Experimental}

For the nonlinear ac conductivity measurements, we prepared
samples of the glasses 0.25 Na$_2$O $\cdot$ 0.096 CaO
$\cdot$ 0.062 Al$_2$O$_3$ $\cdot$ 0.592 SiO$_2$ (NCAS25)
and 0.25 Na$_2$O $\cdot$ 0.75 SiO$_2$ (NS25) with different
thicknesses. The preparation of the NCAS25 glass is described in
Ref. \cite{Murugavel04}. The NS25 glass was prepared by using the
same procedure.

The experimental setup for nonlinear conductivity spectroscopy
and the high-voltage measurement system are described
in detail in Ref. \cite{Murugavel04}. An important aspect of our method is
the utilisation of nonblocking electrodes consisting of highly
conducting NaCl solutions. Thereby, we avoid (i) electrochemical reactions at the
electrode/sample interfaces and (ii) electron or hole injection into the sample.
The measurements were carried out in a frequency range from 10 mHz to
10 kHz with a maximum voltage amplitude of 500 V.

\section{\label{Results}Results}

\begin{figure}
\includegraphics[width=8.6cm]{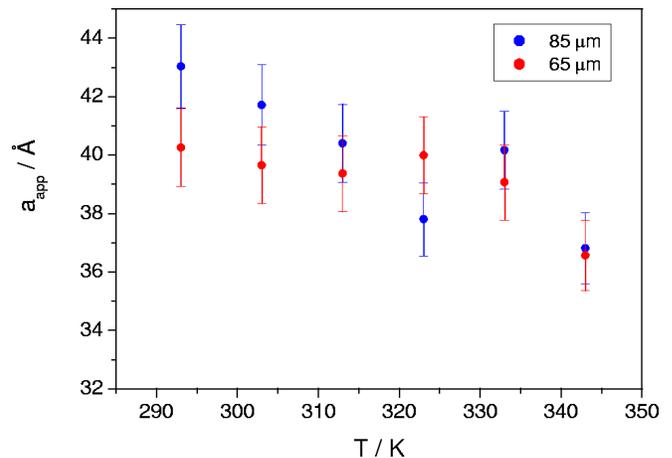}
\caption{\label{fig:a_app} Apparent jump distance of mobile
Na$^+$ ions, $a_{app}$, in NCAS25 glass samples with
different thicknesses.}
\end{figure}

For the NCAS25 glass, the third-order conductivity coefficients
$\sigma_{3,dc}$ were derived from an analysis of the higher
harmonic components of the current density spectrum as described
in Ref. \cite{Murugavel04}. At low frequencies, both the low-field
conductivity $\sigma_1(\nu)$ and the third-order conductivity
coefficient $\sigma_3(\nu)$ exhibit plateaus, the plateau values
being identical to the bulk dc values, $\sigma_{1,dc}$ and
$\sigma_{3,dc}$ \cite{Murugavel04}. From these dc values, the
apparent jump distance $a_{app}$ were calculated using Eq.
(\ref{aappdef}). A careful analysis of the frequency dependence of
the  conductivity data revealed, however, that in the
low-frequency region, sample/electrode interface impedances have a
weak influence on the conductivity. This will be discussed later
in more detail for the NS25 glass where the interfacial impedance
effects are more pronounced. For the determinination of
$\sigma_{1,dc}$ and $\sigma_{3,dc}$ of the NCAS25 glass, the
analysis was limited to a frequency regime where the interfacial
impedance was less than 1\% of the overall impedance. Due to the
narrower frequency regime, the values obtained for $a_{app}$ were
a few percent lower than those published in Ref.
\cite{Murugavel04}. In Fig. \ref{fig:a_app} we show a plot of
$a_{app}$ versus temperature T for two NCAS25 glass samples with
thicknesses 65 $\mu$m and 85 $\mu$m. Within the experimental
error, we do not detect a significant thickness dependence of
$a_{app}$.

\begin{figure}
\includegraphics[width=8.6cm]{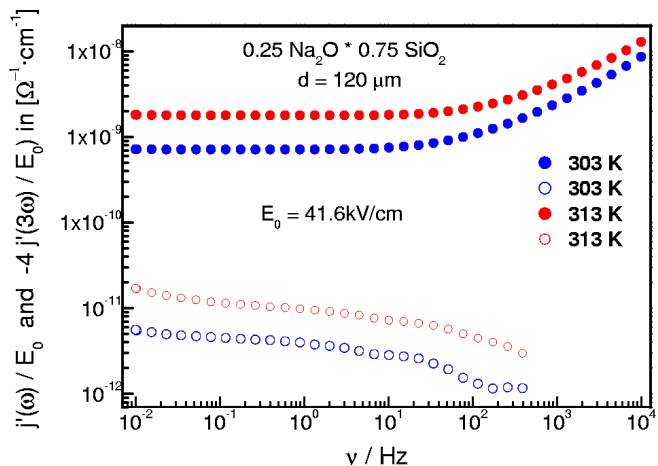}
\caption{\label{fig:ns25_j3omega} Base current density $j'(\nu)$ (closed symbols)
and higher harmonic  current density $j'(3\nu)$ (open symbols),
measured after applying an ac field with
amplitude $E_0$ = 41.6 kV/cm to the NS25 glass. Both Fourier components of
the current density are normalised by the field amplitude $E_0$ and are plotted versus
frequency $\nu$ at two different temperatures, $T$ = 303 K and  313 K.
In addition, the quantity $j'(3\nu)/E_0$ is multiplied by a factor
of -4.}
\end{figure}

In the case of the NS25 glass, we carried out measurements on samples
with thicknesses 83 $\mu$m, 100 $\mu$m and 120 $\mu$m. Unfortunately,
for all samples, the higher harmonic current density $j'(3\nu)$ did not exhibit
well-defined low-frequency plateaus. As an example, we show in Fig.
\ref{fig:ns25_j3omega} results for the base current density $j'(\nu)$
(closed symbols) and for the higher harmonic current density
$j'(3\nu)$, measured after applying an ac field with amplitude
$E_0$ = 41.6 kV/cm to a sample with thickness $d$ = 120 $\mu$m.
Both $j'(\nu)$ and $j'(3\nu)$ are normalised by the field amplitude
$E_0$ and are plotted versus frequency at two different temperatures
$T$ = 303 K and 313 K. While a bulk dc plateau in $j'(\nu)$ is
clearly detectable, the higher harmonic current density $j'(3\nu)$
exhibits a frequency dependence over the entire frequency range.
Therefore, it was not possible to obtain accurate values for
$\sigma_{3,dc}$ of this glass.

\begin{figure}
\includegraphics[width=8.6cm]{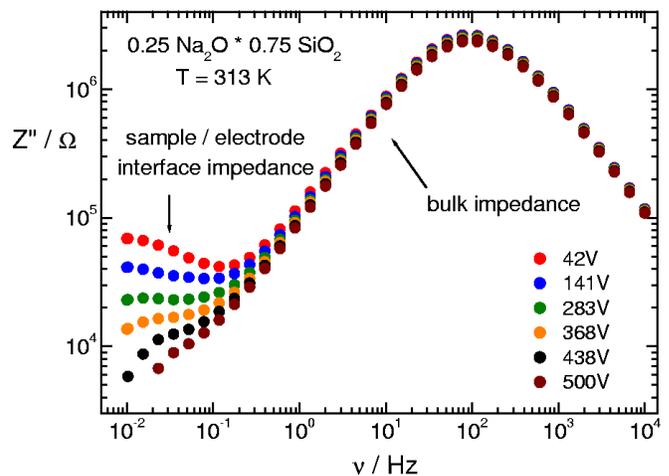}
\newline
\newline
\includegraphics[width=8.6cm]{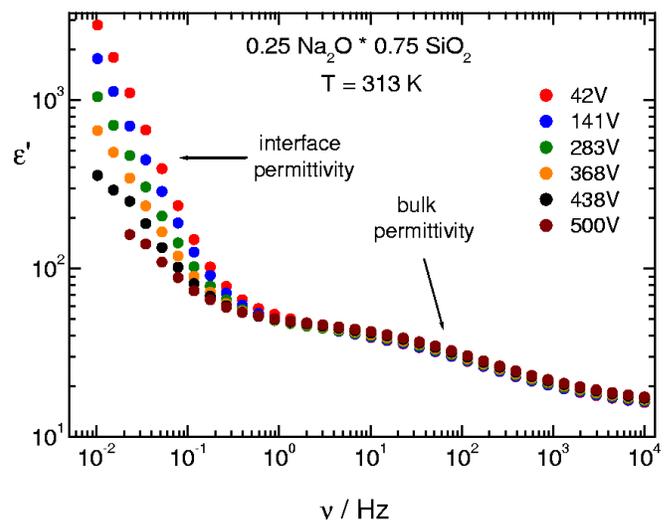}
\caption{\label{fig:ns25_impedance} Frequency dependence of the imaginary
part of the impedance, $Z''$ and of the real part of the permittivity,
$\varepsilon'$, for the
NS25 glass. The data were taken at $T$ = 313 K and different voltage
amplitudes. Above 1 Hz, the spectra are determined by bulk ionic conduction,
while below one 1 Hz, sample/electrode interface impedances govern the
frequency and field dependence of the electrical data. The interfacial
impedances are related to additional barriers for the diffusion of the sodium ions
from the glass samples into the NaCl solution and vice versa.}
\end{figure}

The reason for the absence of well-definded dc plateaus
in $j'(3\nu)$ are most likely electrical impedances at
the sample/electrode interfaces. Although we use highly
conducting liquid NaCl solutions as electrodes, we detect significant
interfacial impedances at low frequencies, which are related to additional
barriers for the diffusion of the sodium ions from the glass samples into
the NaCl solution and vice versa. As an example, we show in Fig.
\ref{fig:ns25_impedance} plots of the imaginary part of the impedance
and of the real part of permittivity versus frequency for the NS25
glass. The data were obtained at $T$ = 313 K and different applied
voltages. At high frequencies, the impedance and permittivity data
are determined by ionic conduction in the bulk and exhibit only
a weak voltage dependence. However, below 1 Hz,
additional impedance contributions from the sample/electrode
interfaces are detected. These interfacial impedances decrease
strongly with increasing voltage. However, even at the highest
applied voltages, interfacial impedance contributions are still present.

\section{\label{Discussion}Discussion}

Although we did not detect a significant thickness dependence of the apparent hopping distance
$a_{app}$ for the NCAS25 glass, our experimental results do not completely
rule out such a thickness dependence. For an unambiguous prove or disprove we will have to study
samples with a broader range of thicknesses. In the case of the NS25 glass,
the thickness was varied by about 50\%, however unfortunately, accurate
$\sigma_{3,dc}$ values for this glass could not be obtained, most likely
due to sample/electrode interface impedance contributions to $j'(3\nu)$.

It is remarkable that the interfacial impedances persist up to voltages
of several hundred volts. A voltage drop of the order of 100 V in an
extremely thin interface region should lead to enormously high electric
fields which should pull the ions over the interfacial barriers. When we assume that
the thickness of the interfacial region is of the order of 10 nm, a voltage drop of the
order of 100 V in this region leads to field strengths of the order
of 100 MV/cm. This is even far above the electrical breakdown strengths
of ion conducting glasses.

From this we conclude that the interfacial regions causing the
interfacial impedance effects must be much thicker than 10 nm. A
possible explanation is the build-up of a resistive surface layer
due to water corrosion. It is thinkable that water diffuses from
the NaCl solution into the glass and and changes the chemical
structure close to the surface by breaking Si-O-Si bonds and
forming Si-O-H groups \cite{Tomozawa85}. The thickness of such
chemically modified surface layers may be much larger than 10 nm
\cite{Tomozawa85}. This assumption would explain the more
pronounced interfacial impedance effects in the case of the NS25
glass. It is well known that the chemical corrosion of simple
alkali silicate glasses is much faster than that of technical
silicate glasses containing additional alkaline-earth oxides. If
the assumption is correct, the interfacial impedances should be
avoidable by using non-aquaous salt solutions, based for instance
on glycerol. This will be the subject of further experiments.


From a theoretical point of view, the observation of large values
of $a_{app}$ has been reproduced by a simple model, describing the
hopping dynamics of a particle (or vacancy) in a disordered model
potential. In the 1D case it is even possible to solve the problem
analytically. This simple and well-studied model contains the
surprising property that $a_{app}$ scales with the thickness.
Identification of $a_{app}\equiv (24 \sigma_{3,dc}/q^2\beta^2
\sigma_{1,dc})^{1/2}$ as some kind of effective hopping distance
was motivated by the solution of the regular model. The scaling
property $a_{app} \propto a N^{1/2}$ for the disordered hopping
model directly implies, however, that this interpretation is {\it
not} justified. Formally, this can be traced back to the fact that
the field-dependence of the current in the disordered case is not
only goverened by the new transition rates $\Gamma_{i,\pm}$ but
also by the modified populations; see Eq. \ref{current}. Rather
for this model the value $a_{app}/(a N^{1/2})$ expresses
information about the nature of the disorder; see Eq.
(\ref{asympt}). Thus, the experimental observation of $a_{app}
\approx 50$\AA \; does not contradict the generally accepted idea
that typical hopping distances in ion conductors should be smaller
than, let's say, 10\AA.

\section{\label{Conclusions}Conclusions}

Combining a detailed experimental analysis with a thorough model
calculation several new aspects have been elucidated. (i) The
large values for $a_{app}$ in thin samples is not related to Joule
heating, but is indeed a reproducible property of the ion
dynamics. (ii) These values increase with decreasing temperature.
(iii) The large values of $a_{app}$ as well as the temperature
dependence can be rationalized by a simple hopping model. (iv) In
general, $a_{app}$ cannot be interpreted as an apparent hopping
distance.

Many new and important questions have emerged from the present
results. (1) What is the real physical interpretation of
$a_{app}$? In particular it would be helpful to relate this
quantity to equilibrium properties of the system
\cite{Dyre89,Dieterich98}. (2) Is is possible to refine the model
to get a limiting value for $a_{app}$? One may speculate that
consideration of interaction effects between different particles
or holes, respectively, may yield such an upper limit. (3) Do
large electric fields also modify the properties of the network
and thus indirectly also the ion dynamics? The latter effect has
not been included in the modelling. In principle this question
could be answered by performing appropriate molecular dynamics
simulations of ion conductors. (4) Are experimental values for
$a_{app}$ different for crystalline materials as compared to
disordered materials? The hopping model, analyzed in this work,
might suggest some difference.

\vspace{0.3cm}
{\bf Appendix}
\vspace{0.3cm}

We consider the covariance of $A_k$ and $A_m$ which can be written
as
\begin{equation}
cov(A_k,A_m) = \sum_{i,j} [ \langle f(k,i)f(m,j) \rangle - \langle
f(k,i) \rangle \langle f(m,j) \rangle ].
\end{equation}
Closer inspection of the $f(k,i)$ (see Eq. (\ref{deff})) shows
that for extreme disorder the first term is particularly large if
$i=j$. This can be seen from the fact that without the additional
exponential factor in Eq. (\ref{deff}) one would strictly have
$\langle f(k,i)f(m,j) \rangle - \langle f(k,i) \rangle \langle
f(m,j) \rangle = 0$ for $i \ne j$. As a consequence one can write
$cov(A_k,A_m)$ as a sum of $N$ terms of similar magnitude such
that $cov(A_k,A_m) \propto N$. Thus the variance of $\sum_m A_m$
which can be written as $\sum_{k,m} cov(A_k,A_m)$ scales like
$N^3$. This has to be compared with the scaling $(\sum_m A_m)^2
\propto N^4$. Thus for large $N$ the standard deviation of $\sum_m
A_m$ is much smaller than the value of $\sum_m A_m$ itself. This
justifies Eq. (\ref{disorder}).

\begin{acknowledgments}
We acknowledge helpful discussions with B. D\"unweg, J.C. Dyre and
R. Friedrich and the support by I. Greger and M. Tsotsalas in the
initial period of the model simulations. Financial support by the
Deutsche Forschungsgemeinschaft and by the Fonds der Chemischen
Industrie is also gratefully acknowledged.
\end{acknowledgments}

\bibliographystyle{apsrev}
\bibliography{mylit}

\end{document}